\documentclass[runningheads]{llncs}
\usepackage[dvipsnames]{xcolor}
\usepackage{graphicx}
\usepackage{amsmath}
\begin{document}
\title{Improved Multi-Shot Diffusion-Weighted MRI with Zero-Shot Self-Supervised Learning Reconstruction}
\titlerunning{Zero-MIRID: Improved Multi-shot Diffusion MRI Reconstruction}
%

\author{Jaejin Cho\inst{1,2}\orcidID{0000-0001-5672-6765} \and
Yohan Jun \inst{1,2}\orcidID{0000-0003-4787-4760} \and
Xiaoqing Wang \inst{1,2}\orcidID{0000-0001-7036-7930} \and
Caique Kobayashi \inst{1,2,3,4}\orcidID{0009-0009-8783-2574} \and
Berkin Bilgic \inst{1,2,5}\orcidID{0000-0002-9080-7865} }
%

\authorrunning{J. Cho et al.}
%

\institute{Athinoula A. Martinos Center for Biomedical Imaging, Charlestown, MA 02129, USA \and
Harvard Medical School, Boston, MA 02115, USA \and
University of São Paulo, São Paulo, 05508-060 State of São Paulo, Brazil \and
Technische Universität München, München, 80333 Bavaria, Germany \and
Harvard-MIT Health Sciences and Technology, Cambridge, MA 02139, USA 
\email{jcho18@mgh.harvard.edu}}

%
\maketitle              
\begin{abstract}
Diffusion MRI is commonly performed using echo-planar imaging (EPI) due to its rapid acquisition time. However, the resolution of diffusion-weighted images is often limited by magnetic field inhomogeneity-related artifacts and blurring induced by $T_2$- and $T_2^{*}$-relaxation effects. To address these limitations, multi-shot EPI (msEPI) combined with parallel imaging techniques is frequently employed. Nevertheless, reconstructing msEPI can be challenging due to phase variation between multiple shots. In this study, we introduce a novel msEPI reconstruction approach called zero-MIRID (zero-shot self-supervised learning of Multi-shot Image Reconstruction for Improved Diffusion MRI). This method jointly reconstructs msEPI data by incorporating deep learning-based image regularization techniques. The network incorporates CNN denoisers in both $k$- and image-spaces, while leveraging virtual coils to enhance image reconstruction conditioning. By employing a self-supervised learning technique and dividing sampled data into three groups, the proposed approach achieves superior results compared to the state-of-the-art parallel imaging method, as demonstrated in an in-vivo experiment.

\end{abstract}
\section{Introduction}
Magnetic resonance imaging (MRI) is widely used for diagnosis and treatment monitoring as it provides structural and physiological information related to disease progression. Diffusion MRI (dMRI) measures molecular diffusion in biological tissues and provides microscopic details of tissue architecture, as molecules interact with many different obstacles while diffusing throughout tissues \cite{le2015diffusion}. However, dMRI requires repeated acquisitions with different diffusion directions. Echo-planar imaging (EPI), which enables fast encoding per imaging slice, is commonly used for dMRI due to its fast acquisition time. However, single-shot (ss-) EPI is susceptible to severe susceptibility-induced geometric distortion and $T_2$- and $T_2^{*}$-induced voxel blurring. These artifacts worsen at higher in-plane resolutions as the time required to acquire each line of $k$-space increases approximately linearly.


Multi-shot (ms-) acquisition is an effective approach to mitigate EPI-related artifacts, which segments $k$-space into multiple portions covered across multiple repetition times (TRs) to reduce the effective echo spacing. However, potential shot-to-shot phase variations across multiple EPI shots can introduce additional artifacts. Recent algorithms, such as low-rank prior methods like low-rank modeling of local $k$-space neighborhoods (LORAKS) \cite{Haldar2014-ov,Haldar2016-ou,kim2017loraks,kim2018loraks,lobos2018navigator,lobos2021robust}, and multi-shot sensitivity encoded diffusion data recovery using structured low-rank matrix completion (MUSSELS) \cite{Mani2017-bo}, have successfully addressed this challenge by jointly reconstructing msEPI images through a low-rank constraint applied across the EPI shots.

In recent years, deep learning has emerged as a promising approach for image reconstruction, offering potential solutions to the challenges of existing techniques, including long reconstruction times, residual artifacts at high acceleration factors, and over-smoothing \cite{hammernik2018learning,eo2018kiki,han2019k}. One notable development is model-based deep learning (MoDL), which leverages an unrolled convolutional neural network (CNN) and a parallel imaging (PI) forward model to denoise and unalias undersampled data \cite{Aggarwal2019-er}. MoDL has also been applied to multi-shot diffusion-weighted echo-planar imaging, known as MoDL-MUSSELS, effectively replacing MUSSELS and significantly reducing reconstruction times while achieving comparable results to state-of-the-art methods \cite{Aggarwal2020-wy}. MoDL-MUSSELS includes CNN denoisers in both image- and $k$-spaces, as recent work has demonstrated that utilizing both domains has yielded improvement in performance based on metrics such as PSNR and SSIM \cite{eo2018kiki}. However, it is worth noting that existing deep learning networks for dMRI have typically been trained in a supervised manner, which requires a significant amount of ground truth images that are not easily acquired in EPI acquisitions.

In contrast, self-supervised learning \cite{Akcakaya2019-lu,Yaman2020-ei,Yaman2022-it} does not rely on external training data and can be used in denoising, reconstruction, quantitative mapping, and other applications. Recent advancements in zero-shot self-supervised learning (ZS-SSL) have demonstrated successful scan-specific network training without any external database \cite{Yaman2022-it}. This approach has shown comparable or superior results to supervised networks. However, in dMRI, where the same volume is repeatedly acquired while changing diffusion directions, ZS-SSL typically requires training separate networks for different directions, which can be impractical.

The virtual coil (VC) approach is a highly effective technique for enhancing the performance of parallel MRI \cite{blaimer2009virtual}, particularly in the case of EPI that utilizes partial Fourier acquisition. VC generates virtual coils by incorporating conjugate symmetric $k$-space signals from actual coils. This integration provides supplementary information for missing data points in $k$-space, further being useful when combined with partial Fourier acquisition. Conceptually, the utilization of VC consistently ensures an image quality equivalent to or exceeds that of the image reconstructed without VC. 

In this study, we propose a novel msEPI reconstruction method called zero-MIRID (zero-shot self-supervised learning of Multi-shot Image Reconstruction for Improved Diffusion MRI). Our method jointly reconstructs msEPI data by incorporating zero-shot self-supervised learning-based image reconstruction. Our key contributions are as follows:

\begin{itemize}
    \item We jointly reconstruct multiple-shot images using self-supervised learning.  
    \item We train one network for all diffusion directions, which accelerates training speed and improves performance.
    \item We used network denoisers in both $k$- and image-space and employed the VC \cite{blaimer2009virtual} to improve the conditioning of the reconstruction.
    \item In the in-vivo experiment, the proposed method demonstrates more robust images and better diffusion metrics than the state-of-art PI technique for dMRI.
    \item To our best knowledge, this study proposes the first self-supervised learning reconstruction for dMRI.
\end{itemize}

Overall, our zero-MIRID method offers a promising approach to enhance msEPI reconstruction in dMRI, providing improved image quality and diffusion metrics through the integration of self-supervised learning techniques.

\section{Method}
\subsection{PI techniques for dMRI}

For msEPI data, SENSE is commonly used for image reconstruction. SENSE individually reconstructs each shot's data using the spatial variation of the coil sensitivity profile. The $m^{th}$ shot image in the $d^{th}$ diffusion direction, $x_{d,m}$, can be reconstructed as follow.

\begin{equation}
x_{d,m} = \underset{x_{d,m}}{\mathrm{argmin}} {\left\|\mathbf{\mathcal{F}}_m\mathbf{C}x_{d,m}-b_{d,m} \right\|}_2^2 
\end{equation}

\noindent where $\mathcal{F}_{m}$ is the undersampled Fourier transform for the $m^{th}$ shot, C is the coil sensitivity map, and $b_{d,m}$ is the acquired $k$-space data of $d^{th}$ direction and $m^{th}$ shot. 

On the other hand, MUSSELS and LORAKS jointly reconstruct multiple-shot images using the low-rank property among msEPI data. The images in the $d^{th}$ diffusion direction can be reconstructed using LORAKS as follows.

\begin{equation}
x_d=\underset{x_d}{\mathrm{argmin}}\sum_{m=0}^{M}{{\left\|\mathbf{\mathcal{F}}_m\mathbf{C} x_{d,m}-b_{d,m} \right\|}_2^2}+\lambda\mathcal{J}(\mathbf{\mathcal{F}}x_d)
\end{equation}

\noindent where $\mathcal{J}$ is the LORAKS regularization. In this work, we utilized S-LORAKS, which employs phase information and $k$-space symmetry \cite{kim2017loraks,kim2018loraks}.

\subsection{Network design}

\begin{figure}[!ht]
\includegraphics[width=\textwidth]{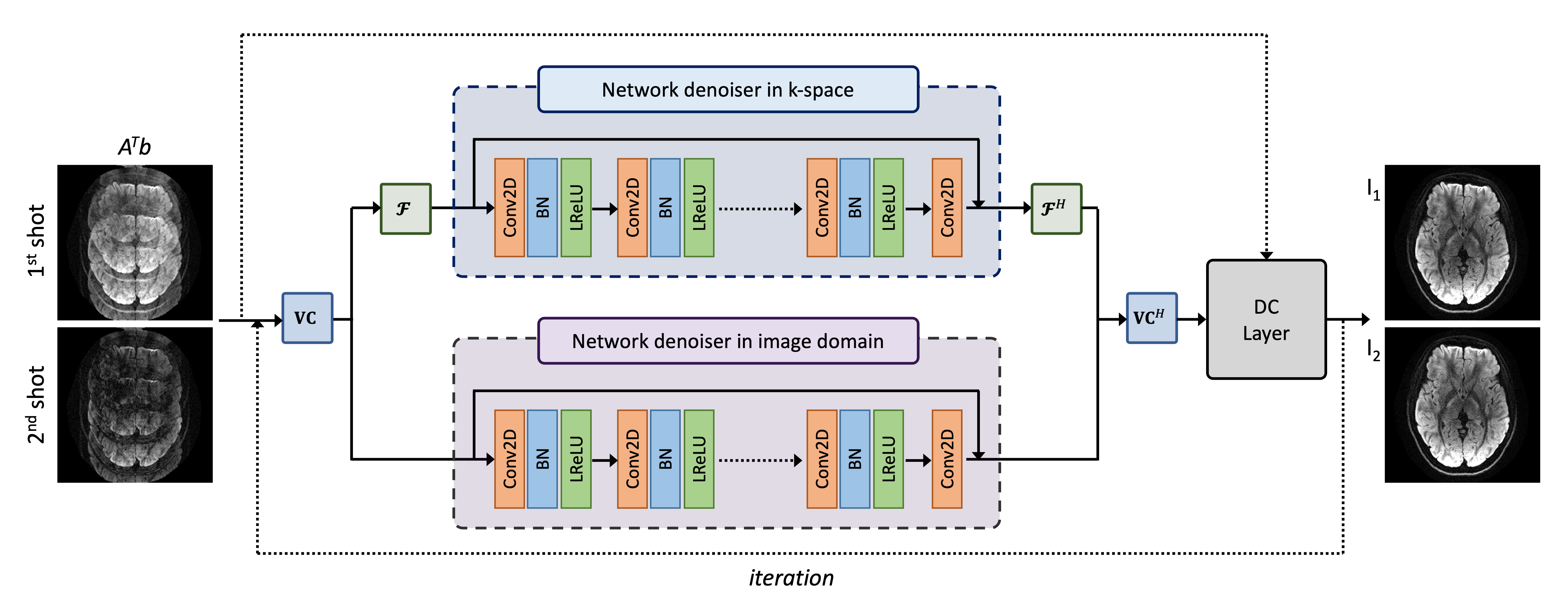}
\caption{The proposed image reconstruction diagram of zero-MIRID. The virtual coil (VC) layer was used to efficiently reconstruct the data accelerated by partial Fourier. The network denoisers in both the $k$-space and image domain were used. The DC layer enforces the consistency between the acquired data and the reconstructed images.} \label{fig1}
\end{figure}

Fig.~\ref{fig1} shows the proposed network diagram of zero-MIRID. The input of the network is $A_m^Tb_d$, where $A_m=\mathcal{F}_mC$. The network consists of two CNNs in the $k$- and image-spaces. The VC was added and removed before and after the denoising CNNs, respectively. The images in the $d^{th}$ diffusion direction can be jointly reconstructed using zero-MIRID as follows.

\begin{equation}
\begin{aligned}
x_d = & \underset{x_d}{\mathrm{argmin}}\sum_{m=0}^{M}{{\left\|\mathbf{\mathcal{F}}_m\mathbf{C}x_{d,m}-b_{d,m} \right\|}_2^2} \\
& + \lambda_1\left\|\mathcal{V}_C^H{N}_i\mathcal{V}_Cx_d\right\|_2^2+\lambda_2\left\|\mathcal{V}_C^H\mathcal{F}^H{N}_k\mathcal{F}\mathcal{V}_Cx_d\right\|_2^2
\end{aligned} 
\end{equation}

\noindent where $\mathcal{V}_C$ is the VC operator, and $N_i$ and $N_k$ are denoising CNNs in the image- and $k$-space, respectively. We define $Nx=x-Dx$, where D is the CNN network, and modified the alternating minimization-based solution in \cite{Aggarwal2020-wy} to get the solutions of equation (3), as follows. 

\begin{equation}
\begin{aligned}
x_{n+1} = & (A^H A + \lambda_1 I +  \lambda_2 I) (A^H b +  \lambda_1 \eta_n +  \lambda_2 \zeta_n) \\
\zeta_{n+1} = & \mathcal{V}^H_C \mathcal{F} ^H D_k \mathcal{F} \mathcal{V}_C x_{n+1} \\
\eta_{n+1} = & \mathcal{V}^H_C D_i \mathcal{V}_C x_{n+1}
\end{aligned}
\end{equation}

\noindent where $n$ is the optimization step (iteration) number, $\eta$ and $\zeta$ is the network denoising terms in $k$- and image-space, and $A=\mathcal{F}C$.

\subsection{Zero-shot self-supervised learning}

\begin{figure}[!ht]
\includegraphics[width=\textwidth]{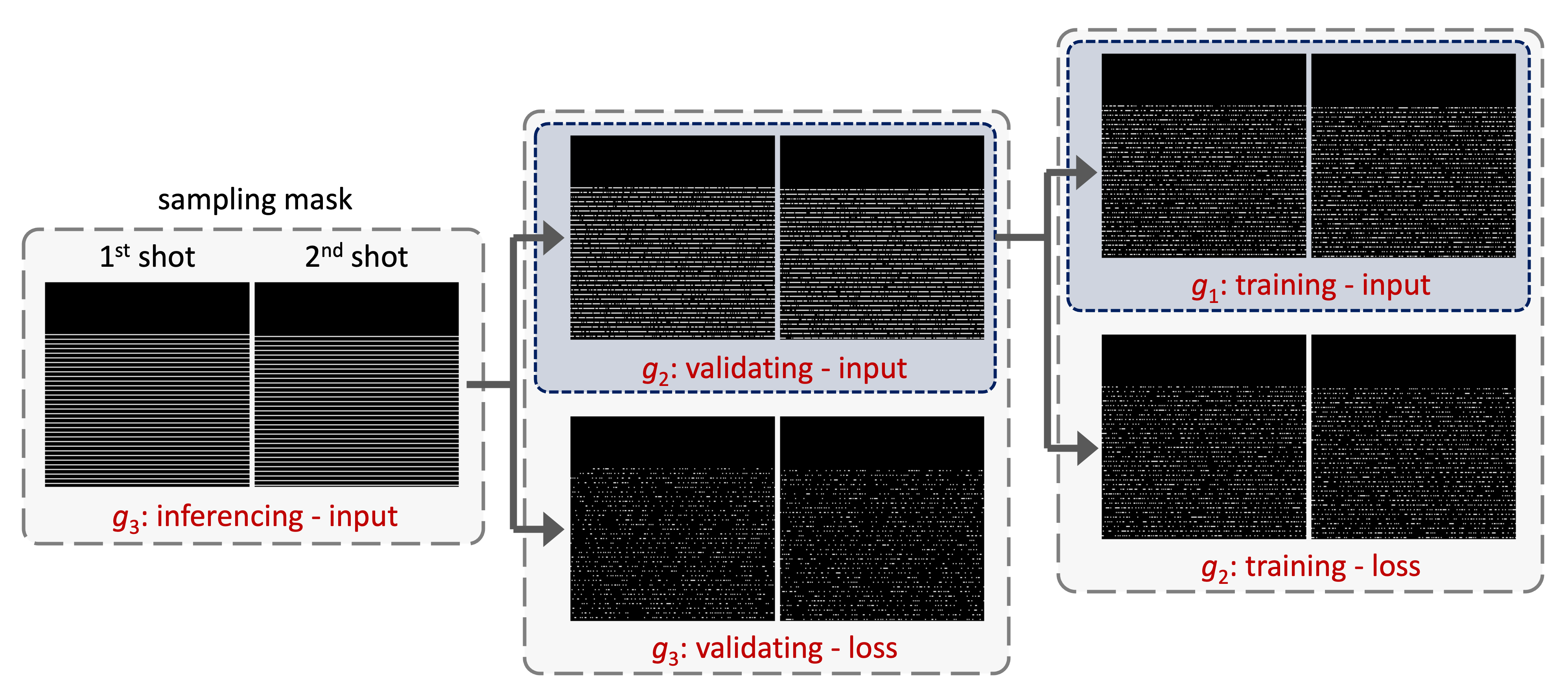}
\caption{The masks used for training, validation, and inference phases. The sampling mask was split into three different masks.} \label{fig2}
\end{figure}

As proposed in the recent ZS-SSL study \cite{Yaman2022-it}, we split the sampling mask into three different groups, as shown in Fig.~\ref{fig2}, where $g_{3}$ is the entire sampling mask and $g_3 \supset g_2 \supset g_1$. In the training phase, $g_{1}$ was used for network input, while $g_{2}$ was used to calculate training losses. In the validating phase, $g_{2}$ was used for network input, while $g_{3}$ was used to calculate validating losses. In the inferencing phase, $g_{3}$ was used for network inputs. The loss in the $d^{th}$ direction in the training phase can be described as follows.

\begin{equation}
\mathcal{L}(g_2 \cdot  b_{d}, \: g_2 \cdot A f(g_1 \cdot  b_{d}; \theta) )
\end{equation}

\noindent where $\mathcal{L}$ is the loss function, $f$ is the zero-MIRID reconstruction, and $\theta$ is the trainable network parameters. Similarly, the loss in the $d^{th}$ direction in the validating phase can be described as follows.

\begin{equation}
\mathcal{L}(g_3 \cdot  b_{d}, \: g_3 \cdot A f(g_2 \cdot  b_{d}; \theta) )
\end{equation}

In this study, we used the normalized root mean square error (NRMSE) and normalized mean absolute error (NMAE) as the loss functions. 

\subsection{Experiment Details}

In-vivo experiments were conducted on a 3T Siemens Prisma system with a 32-channel head coil. For dMRI, we acquired the diffusion-weighted data in 32 different directions using 2-shot EPI, with each shot accelerated by 5-fold (R=5) and employing 75\% Partial Fourier, resulting in 15\% coverage of the $k$-space in each shot relative to a fully-sampled readout. Imaging parameters are; field of view (FOV)=$224\times224\times128$ $mm^{3}$, voxel size =$1\times1\times4$ $mm^{3}$, TR=3.5 s, and effective echo time (TE) =59 ms.

SENSE and S-LORAKS reconstructions were performed with MATLAB R2022a using Intel Xeon 6248R and 512 GB RAM. All neural network implementations were conducted with Python, using the Keras library in Tensorflow 2.4.1. NVIDIA Quadro RTX 8000 (RAM: 48 GB) was used to train, validate, and test the network. The denoising CNNs consist of 16 layers of which the depth is 46. For the 16 layer-CNN, we employed a filter size of 3x3. The depth of our network is 46, resulting in a total of 583,114 trainable parameters. The DC layer takes ten conjugate gradient steps, and the reconstruction block iterates ten times, where the MoDL paper \cite{Aggarwal2019-er} has demonstrated the saturated performance. For training the model, the Adam optimizer is used with a learning rate of 1e-3. Leaky ReLU was used as the activation function. For every diffusion direction, one $g_{2}$ and 50 cases of $g_{1}$ were generated. The ratio of the number of $k$-space points of $g_{3}$:$g_{2}$:$g_{1}$ = 1.00:0.80:0.48. We trained a single network for 32 diffusion directions and used that network to reconstruct all directions. For comparison, we trained two separate networks for the individual reconstruction for each shot (zero-SIRID, single-shot image reconstruction). We used the FSL toolbox for diffusion analysis \cite{smith2004advances,woolrich2009bayesian,jenkinson2012fsl}. To estimate multiple fiber orientations, we used the Bayesian Estimation of Diffusion Parameters Obtained using Sampling Techniques (BEDPOSTX) \cite{behrens2007probabilistic,jbabdi2012model,hernandez2013accelerating}. 

Example data and code can be found in the following link: \\ \url{https://github.com/jaejin-cho/miccai2023}

\begin{figure}[!ht]
\includegraphics[width=\textwidth]{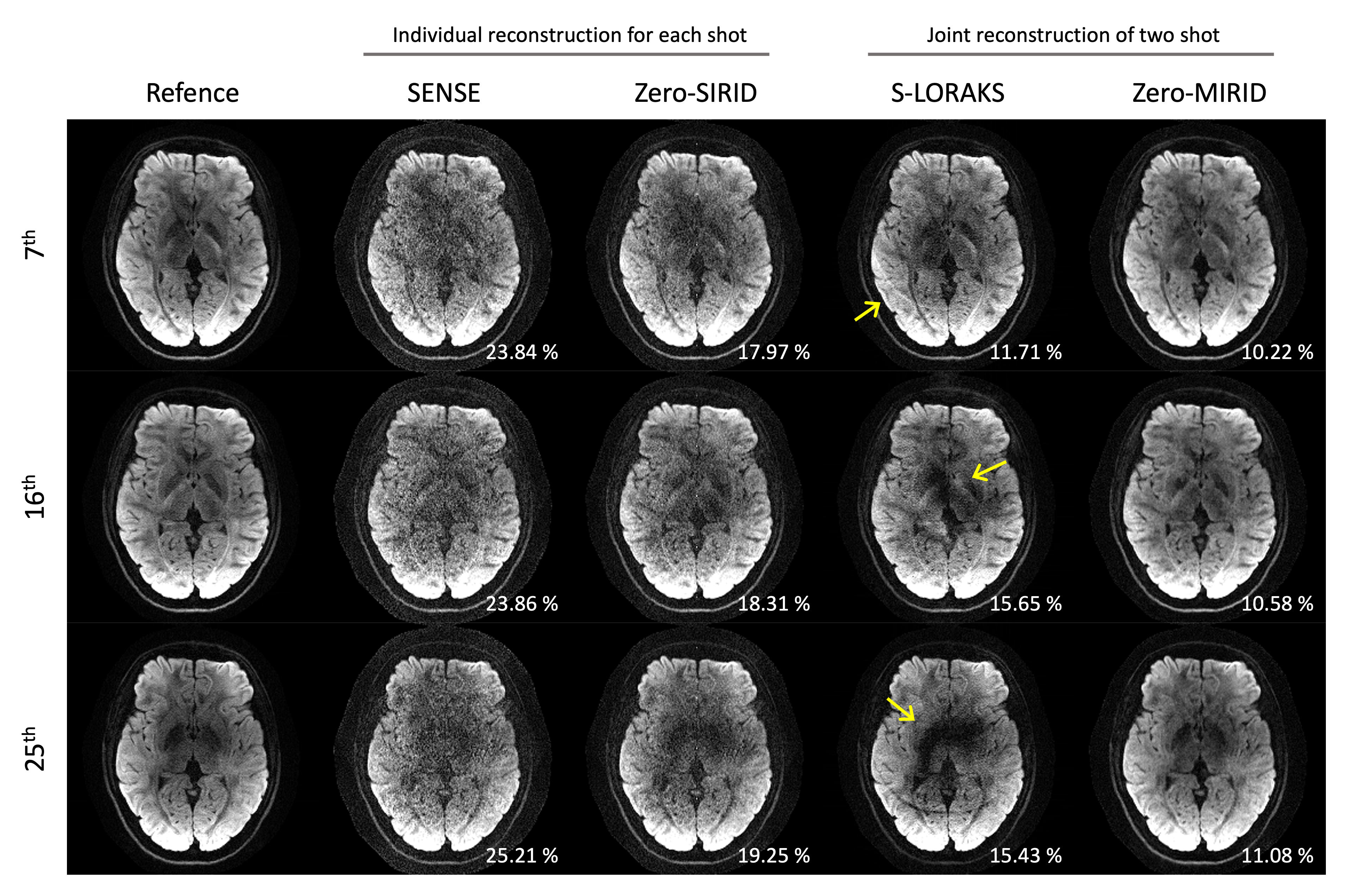}
\caption{The reconstructed diffusion-weighted images at R=5 per shot. Selected diffusion directions were shown. Reference images were obtained from 5-shot EPI data with S-LORAKS reconstruction. SENSE and zero-SIRID individually reconstruct each shot image, whereas S-LORAKS and zero-MIRID jointly reconstruct two shot images. NRMSE was shown at the bottom of each image. } \label{fig3}
\end{figure}

\section{Results}

Fig.~\ref{fig3} the reconstructed diffusion-weighted images at 5-fold acceleration per shot in the selected diffusion directions. The reference images were obtained from 5-shot EPI data that covers complementary $k$-space lines to each other with the S-LORAKS constraint. While SENSE shows severe noise amplification and remaining folding artifacts, zero-SIRID was able to partially mitigate the noise amplification. S-LORAKS jointly reconstructed two shots, considerably reduced noise, and improved the signal-to-noise ratio (SNR). Nonetheless, in the selected diffusion directions, folding artifacts were amplified, and the center of the image shows a dropped signal (pointed by yellow arrows). In contrast, zero-MIRID demonstrated robust image reconstruction even with a high reduction factor per shot. The NRMSE and NAE across the diffusion direction are provided in the supplementary material, demonstrating notable reductions in NRMSE and NMAE when the proposed method is compared to S-LORAKS.

\begin{figure}[!ht]
\includegraphics[width=\textwidth]{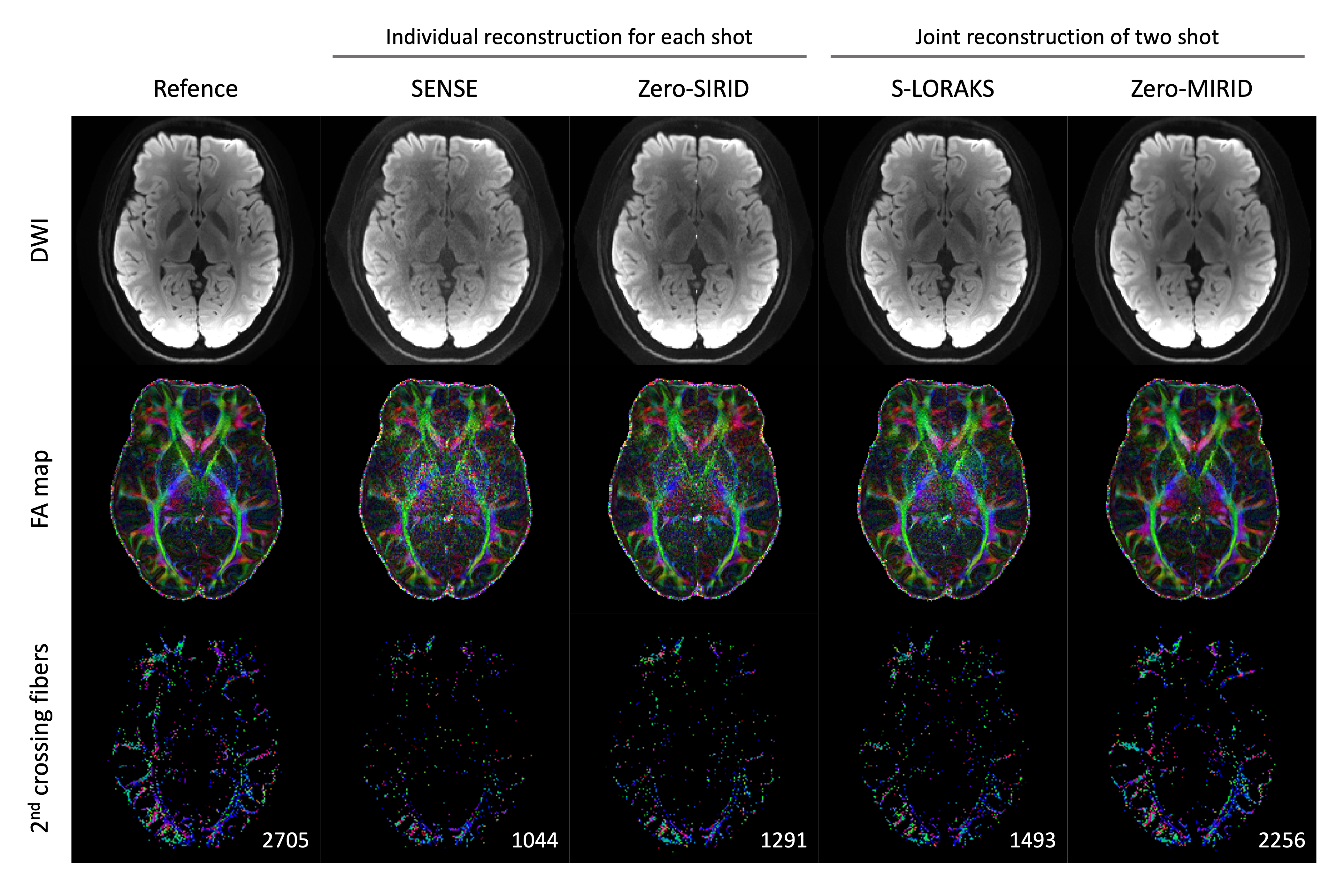}
\caption{Average DWI, FA map, and 2\textsuperscript{nd} crossing fiber image from the reconstructed images in Fig.~\ref{fig3}. The number of 2\textsuperscript{nd} crossing fibers was shown at the bottom of each column.} \label{fig4}
\end{figure}

Fig.~\ref{fig4} presents the average diffusion-weighted image (DWI), fractional anisotropy (FA) map, and 2\textsuperscript{nd} crossing fiber image calculated from the reconstructed images. S-LORAKS and zero-MIRID produced high-fidelity average DWIs, whereas SENSE and zero-SIRID show remaining artifacts. SENSE, zero-SIRID, and S-LORAKS show amplified noise in the center of the FA maps, whereas zero-MIRID effectively mitigated the noise. Furthermore, zero-MIRID well preserved the number of 2\textsuperscript{nd} crossing fibers, often considered a crucial factor in evaluating successful dMRI acquisition \cite{behrens2007probabilistic,jbabdi2012model}.

\section{Discussion and Conclusion}

In this study, we proposed an improved image reconstruction method for msEPI and dMRI in a self-supervised deep learning manner. In-vivo experiment demonstrates the proposed method outperformed S-LORAKS, the state-of-art PI method for dMRI. 

Acquiring reference images of msEPI can be challenging because each shot is typically highly accelerated and shot-to-shot phase variation prevents jointly reconstructing multiple shots efficiently. Advanced PI techniques that jointly reconstruct many EPI shots can improve the PI condition and provide high-fidelity images, but using a PI method may induce bias to that particular method. Therefore, supervised learning might not be an ideal solution for msEPI. On the other hand, self-supervised learning, which does not require reference images, could be a more suitable approach for msEPI. Due to the difficulty in obtaining reliable ground truth data, conventional quantitative metrics such as SSIM and NRMSE may be less reliable for evaluation. In dMRI, FA maps and 2\textsuperscript{nd} crossing fibers could be used for obtaining more suitable metrics.

We trained a single network for all diffusion directions, which improved performance and reduced training time (please see the supplementary material). NRMSE and NMAE were reduced from 14.69\% to 13.61\% and from 15.73\% to 14.41\%, respectively. The training time for the proposed network was 22:30 min per diffusion direction/slice (on GPU). This is expected to be reduced by transfer learning. Inference took approximately 1 second per direction/slice, and 2-shot LORAKS took approximately 20 seconds per direction/slice (on CPU). Since the images are highly similar across diffusion directions, training on the entire diffusion direction has a similar effect to increasing the size of the training database, thereby enhancing network training. Moreover, using a single network for all directions reduces training time compared to training separate networks for each direction, from 40 min to 22:30 min per direction and slice.

As a future work, the simultaneous multi-slice (SMS) technique \cite{setsompop2012blipped}, which is often used for further acceleration, can be easily incorporated into the current network (please see the preliminary images in the supplementary material). At R\textsubscript{sms}=$5\times2$-fold acceleration, NRMSE and NMAE were significantly reduced compared with SENSE, from 22.91\% to 9.07\% and from 26.09\% to 11.12\%, respectively. g-Slider could be a good application as well \cite{setsompop2018high}, because RF-encoded images also have highly similar image features.

\subsubsection{Acknowledgments.} This work was supported by research grants NIH R01 EB028797, R01 EB032378, R03 EB031175, U01 EB025162, P41 EB030006, U01 EB026996, R01 EB017337, U01 DA055353-01, R01 HD100009 and the NVidia Corporation for computing support.

%

\bibliographystyle{splncs04}
\bibliography{paper2502}

\begin{thebibliography}{10}
\providecommand{\url}[1]{\texttt{#1}}
\providecommand{\urlprefix}{URL }
\providecommand{\doi}[1]{https://doi.org/#1}

\bibitem{Aggarwal2019-er}
Aggarwal, H.K., Mani, M.P., Jacob, M.: {MoDL}: {Model-Based} deep learning
  architecture for inverse problems. IEEE Trans. Med. Imaging  \textbf{38}(2),
  394--405 (Feb 2019)

\bibitem{Aggarwal2020-wy}
Aggarwal, H.K., Mani, M.P., Jacob, M.: {MoDL-MUSSELS}: {Model-Based} deep
  learning for multishot {Sensitivity-Encoded} diffusion {MRI}. IEEE Trans.
  Med. Imaging  \textbf{39}(4),  1268--1277 (Apr 2020)

\bibitem{Akcakaya2019-lu}
Ak{\c c}akaya, M., Moeller, S., Weing{\"a}rtner, S., U{\u g}urbil, K.:
  Scan-specific robust artificial-neural-networks for k-space interpolation
  ({RAKI}) reconstruction: Database-free deep learning for fast imaging. Magn.
  Reson. Med.  \textbf{81}(1),  439--453 (Jan 2019)

\bibitem{behrens2007probabilistic}
Behrens, T.E., Berg, H.J., Jbabdi, S., Rushworth, M.F., Woolrich, M.W.:
  Probabilistic diffusion tractography with multiple fibre orientations: What
  can we gain? neuroimage  \textbf{34}(1),  144--155 (2007)

\bibitem{blaimer2009virtual}
Blaimer, M., Gutberlet, M., Kellman, P., Breuer, F.A., K{\"o}stler, H.,
  Griswold, M.A.: Virtual coil concept for improved parallel mri employing
  conjugate symmetric signals. Magnetic Resonance in Medicine: An Official
  Journal of the International Society for Magnetic Resonance in Medicine
  \textbf{61}(1),  93--102 (2009)

\bibitem{eo2018kiki}
Eo, T., Jun, Y., Kim, T., Jang, J., Lee, H.J., Hwang, D.: Kiki-net:
  cross-domain convolutional neural networks for reconstructing undersampled
  magnetic resonance images. Magnetic resonance in medicine  \textbf{80}(5),
  2188--2201 (2018)

\bibitem{Haldar2014-ov}
Haldar, J.P.: Low-rank modeling of local k-space neighborhoods ({LORAKS}) for
  constrained {MRI}. IEEE Trans. Med. Imaging  \textbf{33}(3),  668--681 (Mar
  2014)

\bibitem{Haldar2016-ou}
Haldar, J.P., Zhuo, J.: {P-LORAKS}: Low-rank modeling of local k-space
  neighborhoods with parallel imaging data. Magn. Reson. Med.  \textbf{75}(4),
  1499--1514 (Apr 2016)

\bibitem{hammernik2018learning}
Hammernik, K., Klatzer, T., Kobler, E., Recht, M.P., Sodickson, D.K., Pock, T.,
  Knoll, F.: Learning a variational network for reconstruction of accelerated
  mri data. Magnetic resonance in medicine  \textbf{79}(6),  3055--3071 (2018)

\bibitem{han2019k}
Han, Y., Sunwoo, L., Ye, J.C.: $k$-space deep learning for accelerated mri.
  IEEE transactions on medical imaging  \textbf{39}(2),  377--386 (2019)

\bibitem{hernandez2013accelerating}
Hern{\'a}ndez, M., Guerrero, G.D., Cecilia, J.M., Garc{\'\i}a, J.M., Inuggi,
  A., Jbabdi, S., Behrens, T.E., Sotiropoulos, S.N.: Accelerating fibre
  orientation estimation from diffusion weighted magnetic resonance imaging
  using gpus. PloS one  \textbf{8}(4),  e61892 (2013)

\bibitem{jbabdi2012model}
Jbabdi, S., Sotiropoulos, S.N., Savio, A.M., Gra{\~n}a, M., Behrens, T.E.:
  Model-based analysis of multishell diffusion mr data for tractography: How to
  get over fitting problems. Magnetic resonance in medicine  \textbf{68}(6),
  1846--1855 (2012)

\bibitem{jenkinson2012fsl}
Jenkinson, M., Beckmann, C.F., Behrens, T.E., Woolrich, M.W., Smith, S.M.: Fsl.
  Neuroimage  \textbf{62}(2),  782--790 (2012)

\bibitem{kim2017loraks}
Kim, T.H., Setsompop, K., Haldar, J.P.: Loraks makes better sense:
  phase-constrained partial fourier sense reconstruction without phase
  calibration. Magnetic resonance in medicine  \textbf{77}(3),  1021--1035
  (2017)

\bibitem{kim2018loraks}
Kim, T., Haldar, J.: Loraks software version 2.0: Faster implementation and
  enhanced capabilities. University of Southern California, Los Angeles, CA,
  Tech. Rep. USC-SIPI-443  (2018)

\bibitem{le2015diffusion}
Le~Bihan, D., Iima, M.: Diffusion magnetic resonance imaging: what water tells
  us about biological tissues. PLoS biology  \textbf{13}(7),  e1002203 (2015)

\bibitem{lobos2021robust}
Lobos, R.A., Hoge, W.S., Javed, A., Liao, C., Setsompop, K., Nayak, K.S.,
  Haldar, J.P.: Robust autocalibrated structured low-rank epi ghost correction.
  Magnetic resonance in Medicine  \textbf{85}(6),  3403--3419 (2021)

\bibitem{lobos2018navigator}
Lobos, R.A., Kim, T.H., Hoge, W.S., Haldar, J.P.: Navigator-free epi ghost
  correction with structured low-rank matrix models: New theory and methods.
  IEEE transactions on medical imaging  \textbf{37}(11),  2390--2402 (2018)

\bibitem{Mani2017-bo}
Mani, M., Jacob, M., Kelley, D., Magnotta, V.: Multi-shot sensitivity-encoded
  diffusion data recovery using structured low-rank matrix completion
  ({MUSSELS}). Magn. Reson. Med.  \textbf{78}(2),  494--507 (Aug 2017)

\bibitem{setsompop2018high}
Setsompop, K., Fan, Q., Stockmann, J., Bilgic, B., Huang, S., Cauley, S.F.,
  Nummenmaa, A., Wang, F., Rathi, Y., Witzel, T., et~al.: High-resolution in
  vivo diffusion imaging of the human brain with generalized slice dithered
  enhanced resolution: Simultaneous multislice (g s lider-sms). Magnetic
  resonance in medicine  \textbf{79}(1),  141--151 (2018)

\bibitem{setsompop2012blipped}
Setsompop, K., Gagoski, B.A., Polimeni, J.R., Witzel, T., Wedeen, V.J., Wald,
  L.L.: Blipped-controlled aliasing in parallel imaging for simultaneous
  multislice echo planar imaging with reduced g-factor penalty. Magnetic
  resonance in medicine  \textbf{67}(5),  1210--1224 (2012)

\bibitem{smith2004advances}
Smith, S.M., Jenkinson, M., Woolrich, M.W., Beckmann, C.F., Behrens, T.E.,
  Johansen-Berg, H., Bannister, P.R., De~Luca, M., Drobnjak, I., Flitney, D.E.,
  et~al.: Advances in functional and structural mr image analysis and
  implementation as fsl. Neuroimage  \textbf{23},  S208--S219 (2004)

\bibitem{woolrich2009bayesian}
Woolrich, M.W., Jbabdi, S., Patenaude, B., Chappell, M., Makni, S., Behrens,
  T., Beckmann, C., Jenkinson, M., Smith, S.M.: Bayesian analysis of
  neuroimaging data in fsl. Neuroimage  \textbf{45}(1),  S173--S186 (2009)

\bibitem{Yaman2022-it}
Yaman, B., Hosseini, S.A.H., Akcakaya, M.: {Zero-Shot} {Self-Supervised}
  learning for {MRI} reconstruction. In: International Conference on Learning
  Representations (2022)

\bibitem{Yaman2020-ei}
Yaman, B., Hosseini, S.A.H., Moeller, S., Ellermann, J., U{\u g}urbil, K.,
  Ak{\c c}akaya, M.: Self-supervised learning of physics-guided reconstruction
  neural networks without fully sampled reference data. Magn. Reson. Med.
  \textbf{84}(6),  3172--3191 (Dec 2020)

\end{thebibliography}

\end{document}